# $^3$He-rich SEP Events Observed by STEREO-A

R. Bučík[*], U. Mall[*], A. Korth[*], G. M. Mason[¶] and R. Gómez-Herrero[‡]

[*]*Max-Planck-Institut für Sonnensystemforschung, Max-Planck-Str. 2, 37191 Katlenburg-Lindau, Germany*
[¶]*Applied Physics Laboratory, Johns Hopkins University, Laurel, MD 20723, USA*
[‡]*Space Research Group, University of Alcalá, 28871 Alcalá de Henares, Spain*

**Abstract.** Using the SIT (Suprathermal Ion Telescope) instrument on STEREO-A we have examined the abundance of the rare isotope $^3$He during the rising activity phase of solar cycle 24 between January 2010 and December 2011. We have identified six solar energetic particle (SEP) events with enormous abundance enhancements of $^3$He ($^3$He/$^4$He >1). The events were short lasting, typically ~ 0.5 – 1 day and most of them occurred in association with high-speed solar wind streams and corotating interaction regions. With one exception the events were not associated with ~ 100 keV solar electron intensity increases. The events showed also enhanced NeS/O and Fe/O ratios. The solar images indicate that the events were generally associated with the active regions located near a coronal hole.

## INTRODUCTION

$^3$He-rich (or impulsive) solar energetic particle (SEP) events are characterized by $^3$He/$^4$He ratios enhanced up to 4 orders of magnitude above coronal values. Heavy ions such as Ne and Fe are also enhanced in these events. $^3$He-rich SEP events originate from the localized flaring active regions and therefore magnetic connection to the source is required for observations of these events [1]. Recently, it has been shown that some of these events may be observed over much broader longitude ranges than previously thought [2].

In this study we examine abundance of $^3$He during a 2-year period during the rising phase of solar cycle 24. We focus on $^3$He-rich SEP events where the $^3$He isotope is dominant over $^4$He. Although the events with such enrichments are not unusual they have not been separately investigated. The purpose of this paper is to identify features, common for SEP events with this enormous $^3$He abundance.

## OBSERVATIONS

The two STEREO spacecraft are in a heliocentric orbit at ~ 1 AU, STEREO-A preceding the Earth and STEREO-B trailing behind. Because of its better $^3$He resolution the measurements presented here were made with the Suprathermal Ion Telescope (SIT) on the STEREO-A [3]. The SIT is a time-of-flight mass spectrometer which measures He-Fe from 20 keV/n to a few MeV/n. In our study we also make use of the He measurements made by the ACE/ULEIS [4], energetic electron observations obtained by the SEPT [5], solar wind measurements made by the PLASTIC [6] and by the magnetometer on STEREO-A [7]. In addition, we use STEREO-A/SECCHI solar images [8].

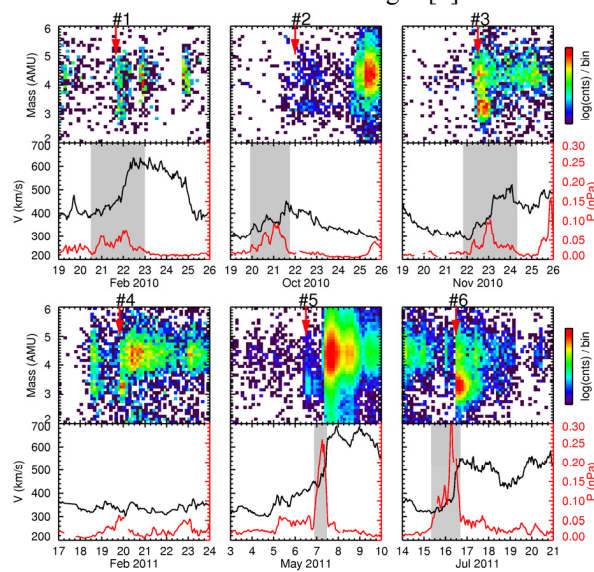

**FIGURE 1.** Upper panels: mass spectrograms at 0.2–0.5 MeV/n. Lower panels: solar wind speed $V$ and total pressure $P$ (right scale). Red arrows mark the beginning of the $^3$He-rich events investigated in this survey. Shaded regions mark the time intervals of CIRs.

**TABLE 1.** Characteristics of the STEREO-A $^3$He-rich SEP events

| Event number | 1 | 2 | 3 | 4 | 5 | 6 |
|---|---|---|---|---|---|---|
| Start day | 2010 Feb 21 | 2010 Oct 21 | 2010 Nov 22 | 2011 Feb 19 | 2011 May 6 | 2011 Jul 16 |
| $^3$He-rich interval (doy) | 52.6-53.1 | 294.8-295.9 | 326.5-327.1 | 50.6-51.1 | 126.4-127.1 | 197.4-198.3 |
| $^3$He/$^4$He (386 keV/n) | 2.60 ± 1.04 | 1.48 ± 0.41 | 1.44 ± 0.19 | 1.40 ± 0.22 | 4.12 ± 0.67 | 4.08 ± 0.29 |
| NeS/O (386 keV/n) | 1.82 ± 0.68 | 1.00 ± 0.82 | 1.18 ± 0.48 | 0.80 ± 0.54 | 0.86 ± 0.48 | 1.48 ± 0.37 |
| Fe/O (386 keV/n) | 1.00 ± 0.43 | 2.67 ± 1.81 | 0.91 ± 0.40 | 0.80 ± 0.54 | 1.43 ± 0.70 | 1.19 ± 0.31 |
| Energetic electrons [a] | Y | N | N | N | N | N |
| Velocity dispersion [b] | ? | N | Y | N | ? | ? |
| Connect. longitude [c] | W51 | W57 | W63 | W69 | W55 | W59 |
| CIR/high-speed stream | Y | Y | Y | N | Y | Y |
| STEREO-A-Earth co-rotation delay (days) [d] | 4.9 | 6.3 | 6.4 | 6.6 | 7.0 | 7.5 |
| ACE $^3$He-rich interval [e] | 45.0-48.0 | … | 318.5-322.5 | … | 119.7-120.3 | 188.5-192.0 |
| ACE CIR interval [f] | … | … | 318.0-319.0 | … | 119.5-120.0 | 190.0-191.0 |

[a] using STEREO-A/SEPT sensor pointing sunward in the energy range 60-155 keV
[b] events #1 and #5 show some indication in the spectrogram of ion energy *vs.* arrival time and event #6 in the He time-intensity profiles
[c] determined form Archimedean spiral angle using STEREO-A/PLASTIC 1hr solar wind speed at event start time
[d] determined from Carrington rotation period 27.3 days and STEREO-A – Earth separation angle at event start time
[e] using ULEIS > 0.5 MeV/n mass spectrograms; the $^3$He enrichment measured by ACE between days 45.0 – 48.0 in 2010 started ~3 days earlier as expected from the corotation delay and therefore probably does not correspond to event #1
[f] using solar wind speed and magnetic field measurements from SWEPAM and MAG instruments on ACE, respectively

In the interval January 2010 – December 2011 we identified six periods with $^3$He/$^4$He > 1 in the 0.320-0.452 MeV/n energy range. These $^3$He-rich periods are shown in Fig. 1. The upper panels present mass spectrograms for the energy range 0.2 – 0.5 MeV/n with the color scale indicating the logarithm of the number of counts in 3 hr x 0.1 amu bins. The lower panels show hourly averages of solar wind speed $V$ (left scale) and the total pressure $P$ – sum of magnetic and plasma proton pressures (right scale). Shaded regions mark the time intervals of corotating interaction regions (CIRs) obtained from a list compiled by the STEREO magnetometer team at the University of California Los Angeles (http://www-ssc.igpp.ucla.edu/forms/stereo/stereo_level_3.html).
Basic characteristics of the investigated events are listed in Tab.1. The event-integrated elemental ratios in the table are shown with statistical errors.

Figure 1 shows that events #1, 2, 3, 5, 6 occurred in the vicinity of the CIRs and the high speed solar wind. Events #5 and #6 have markedly higher peak pressures than the other events. Event #2 is followed by the SEP event on 24 October 2010 with small $^3$He enrichment at the beginning; event #3 is followed by weak CIR event; event #4 is immediately followed by the gradual SEP event. We note that event #4 occurred during the flat phase of the solar wind speed. The $^3$He-rich period on 18 February 2011 (preceding event #4) has a $^3$He/$^4$He ratio below our selection threshold. Event #5 is followed by the strong CIR event. Event #6 started near the trailing edge of the CIR, which is often observed in CIR events (see for example the CIR event after event #5). This raises the question whether the CIR might have played a role in intensifying this $^3$He-rich event.

All investigated events show Fe/O enhanced by ~2-8 and NeS/O by ~1.2-3 when compared to gradual SEP events [9]. In events # 1, 3, 4, 5, 6 the Fe/O is close to the average Fe/O ratio in other impulsive SEP (ISEP) events [10]. Event #2 shows the Fe/O more than two times enhanced compared to ISEP events. All events show the NeS/O close to the average NeS/O ratio in other ISEP events [10].

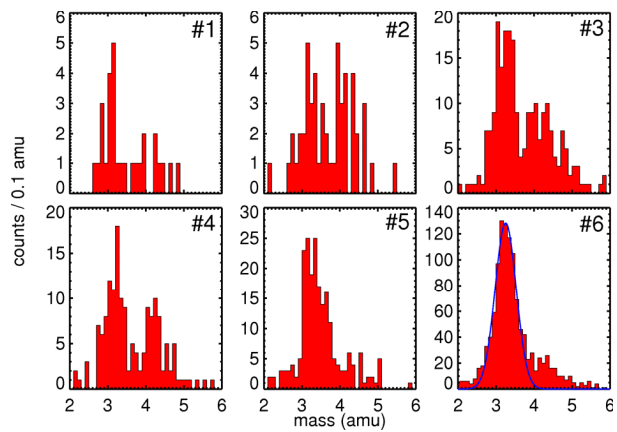

**FIGURE 2.** He mass histograms in the energy range 0.320-0.452 MeV/n for six events in this survey. Blue curve in histogram #6 is a Gaussian fit to the $^3$He peak.

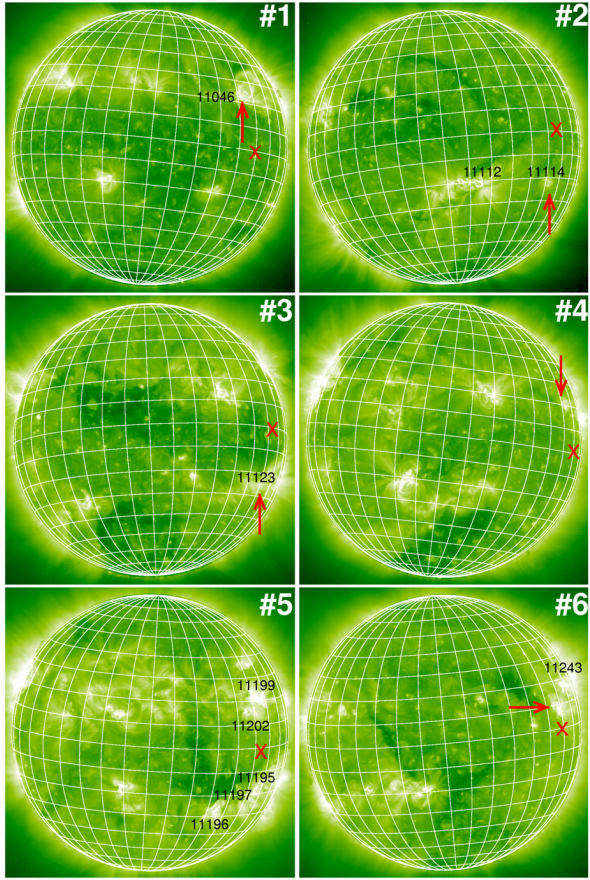

**FIGURE 3.** STEREO-A/SECCHI 195 Å images at start times of the events. Red crosses and arrows are described in the text.

The histograms shown in Fig. 2 have been used for calculation of the $^3$He/$^4$He ratios in Tab.1. The blue curve in histogram #6 is a Gaussian fit to the $^3$He peak. The 3σ upper limit for this fit is at 4.1 amu. Based on the SIT-A helium histograms of 14 CIR events in 2010, which are almost all $^4$He, we found ~36 % of the $^4$He counts below 4.1 amu. Using this value we estimated spillover of the $^4$He to $^3$He and also total $^4$He peak by summing the counts above 4.1 amu for each event. The histograms can provide a guess about He fluences. Strikingly high $^3$He fluence is seen in event #6.

Figure 3 shows STEREO-A/SECCHI EUVI 195 Å images at start times of the events. Red crosses mark approximate STEREO-A connection longitude (see Tab.1). Red arrows point to the isolated active regions which are located near the connection longitude and therefore might be the sources of the $^3$He emissions for our events. The active region (AR) 11046 (panel #1) produced EUV coronal jets with temporally coincident electron spikes on 22 February 2010 [11]. In panel #2 the AR 11112 was responsible for the bigger 24 October 2010 SEP event (see Fig. 1).

Five-minute cadence 195 Å images showed clear brightening at the marked ARs near the predicted particle injection times at the Sun for events #1, 2, and 4. We note that 386 keV/n ions would propagate scatter-free (with zero pitch angle) along the Archimedean spiral from the Sun to 1 AU in ~5.5 hours. Since events #2 and 4 show no velocity dispersion it is likely that the observed $^3$He ions were released earlier than the extrapolated times. The supposed source for event #3 (AR 11123) and #6 did not show obvious brightening near particle injection times. Instead, small bright spots temporarily appeared in the nearby dark areas. The source for event #5 is even less clear. In addition to the weak brightening in the AR 11197 the northern hemisphere ARs 11199 and 11202 showed material ejections around estimated particle release times.

Figure 4 shows potential-field source-surface (PFSS) model field lines that are open to the heliosphere at the ecliptic plane (http://gong.nso.edu/) near the start times of the events. Also shown are the associated magnetogram, the polarity inversion line and coronal holes, which are represented by dots. Positive/negative polarity coronal holes and field lines are indicated by green/red. Black crosses mark STEREO-A connection longitude. The figure indicates that source regions selected in EUVI images contain open field to the ecliptic in events # 1, 3, 4, 6. PFSS mapping shows that STEREO-A is likely connected via positive field to the AR 11112 in event #2. Possible candidates for event #5 have also open field to the ecliptic. We examined the magnetic field polarity at 1 AU and found that this matched the polarity expected from PFSS in all cases.

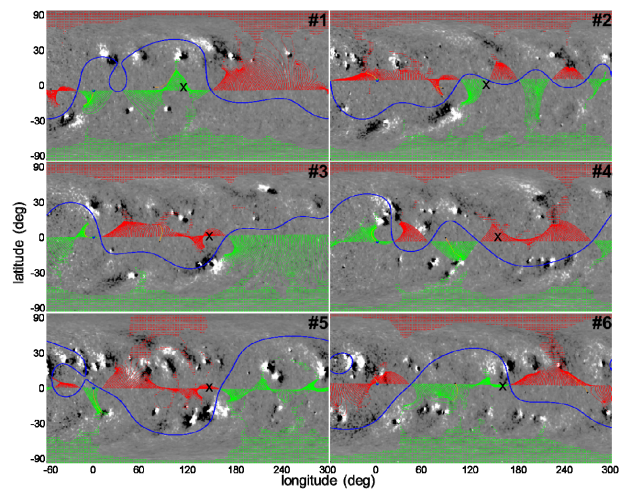

**FIGURE 4.** PFSS ecliptic field lines. Green/red indicates positive/negative open field. Black crosses mark STEREO-A connection longitude.

## DISCUSSION AND CONCLUSION

EUVI solar images and PFSS extrapolations indicate that the events are generally associated with the active regions located near the coronal hole. Such regions as a source of $^3$He-rich SEPs were identified in the recent observations [12]. The EUV jets, which are considered to be signatures of the reconnection between active region and coronal hole field lines, were regularly seen in $^3$He-rich SEP events [12, 13]. Jet-like ejections have also been reported in the source for one event in this survey (#1).

High-speed streams and CIRs were often present in the investigated events. The appearance of the $^3$He-rich events at the onset of high-speed streams has been reported in the previous survey [14]. It has been thought that flare particles have a higher probability of escape into the interplanetary space on open-field-line regions in the solar wind streams. We found that while all pure $^3$He-rich events (#1, 2, 3, 5, 6) were associated with high-speed solar wind the two $^3$He-rich events which are followed by a gradual event (event #4 and the 24 October 2010 event) occurred during the flat phase of the slow solar wind.

The two events in our study (#5 and #6) with the highest $^3$He enrichment and fluence were accompanied by the CIRs with the highest compression. Interpreting multiday $^3$He-rich periods, it has been recently suggested [15] that CIRs can temporarily trap and re-accelerate SEPs. This may be a plausible explanation for enormous $^3$He enrichments in these two events. It has been also concluded that the magnetic connection to the solar source plays an important role in determining of the observed $^3$He enrichment [12] and $^3$He fluence [16].

The combination of the $^3$He observations from ACE and STEREO-A shows that at least three of the six events (#3, 5, 6) were re-current, i.e. the event was observed on STEREO-A with delay corresponding to the corotation time between ACE and STEREO-A. This might happen either because the regions of $^3$He emission were long-active (~ seven days) or the particles were confined in the CIRs. The first possibility is likely true for event #3 where the clear velocity dispersion suggests new injection when the responsible region rotated to the STEREO-A connection point.

The abundances in the investigated events suggest that the enrichments in the heavier ions and $^3$He are not related. For example, event #2 shows Fe/O greatly enhanced while the $^3$He/$^4$He ratio has one of the lowest values. This feature has been known from earlier observations [17] and indicates that different mechanisms or different solar regions are involved in enrichments of these nuclei.

Energetic electrons, the signature associated with $^3$He-rich SEPs, were generally not seen in the investigated events. It could be that the electron emissions were dominated by lower energies or they were simply missing. Thus the processes responsible for huge $^3$He enrichments are likely different from those operating in flare electron acceleration.

In conclusion, we found that the $^3$He-rich events with $^3$He/$^4$He > 1 examined in this survey are usually associated with CIRs and solar wind streams, but not with energetic electrons. Their associated solar sources have a tendency to be located near a coronal hole. The observations suggest that CIRs probably play a role in the $^3$He enrichments in addition to other factors discussed in past studies.

## ACKNOWLEDGMENTS

This work was supported by the Bundesministerium für Wirtschaft under grant 50 OC 0904. The work at JHU/APL was supported by NASA under contract SA4889-26309 from the University of California Berkeley and by NASA grant 44A-1089749. This work utilizes data obtained by the Global Oscillation Network Group (GONG) program, managed by the National Solar Observatory. We thank the ACE SWEPAM/MAG instrument teams and the ACE Science Center for providing the ACE data. We thank the referee for comments that have improved the manuscript.